# Realizing a New Research Agenda for Writing-to-Learn:

## Embedding Process in Context

## Manuscript Submission to Educational Researcher

## Corresponding Author:

Lisa M. Hermsen
Department of English
Rochester Institute of Technology
Rochester, NY 14623
Office Phone: (585) 475-4553
Fax: (585) 475-5097
lmhgsl@rit.edu

## Author:

Scott V. Franklin
Department of Physics
Rochester Institute of Technology
Rochester, NY 14623
Office Phone: (585) 475-2536 _
Lab Phone: (585) 475-2577 _
Fax: (585) 475-2536
_svfsps@rit.edu

# Realizing a New Research Agenda for Writing-to-Learn:

**Embedding Process in Context**

ABSTRACT

Writing-to-learn initiatives such as Writing Across the Curriculum or Writing in the Disciplines occupy the center of writing programs nationwide. Nevertheless, research to support the core of the writing-to-learn philosophy–that the writing process can facilitate content learning–is, at best, inconclusive. Calls for additional research have noted the importance of either additional cognitive analyses or new contextual research methods. This paper argues for a unique research agenda that would embed cognitive processes into disciplinary contexts and thus provide a layered, multi-modal research approach into writing and learning.

# Realizing a New Research Agenda for Writing-to-Learn:
**Embedding Process in Context**

Since the early 1970s Writing Across the Curriculum (WAC) and Writing in the Disciplines (WID) initiatives have theorized a link between language and learning, positing writing as a higher order tool for content learning ("write-to-learn", e.g. Applebee, 1985; Emig 1977; Fulwiler & Young, 1982; Knoblauch & Brannon, 1983; McLeod, 1989; Odell, 1980). Other initiatives promoting WID theorize that writing within a specific discipline introduces students to that discipline's epistemology (Herrington, 1981; Kaufer & Young, 1993; Kiefer, 1990; Russell, 1991), thus facilitating content learning. WAC and WID models have achieved institutional endorsement and academic acceptance because they promote intuitively accurate practices that persist as theoretically sound. Nevertheless, research into the WAC and WID models' specific claims has yet to develop so as to provide conclusive supporting evidence. This paper argues for an empirical educational research design that treats writing as a socio-cognitive phenomenon within a complex understanding of context. This multimodal design, we argue, is unique in testing the fundamental assumptions of WAC and WID connecting learning with discipline-specific writing behaviors and patterns.

The theory and practice of WAC and WID has burgeoned as the major strategies in teaching academic writing, often resulting in "writing intensive" courses even in the "hard sciences." The initiatives have nevertheless undergone thin assessment and can claim very little supporting research. Research into the theory and practice of WAC and WID initiatives has been limited to local, classroom-based student writing assessments and/or program-specific assessments (Bazerman, et al, 2005; Bazerman and Russell, 1994; Walvoord. 1992; McLeod, 1988; Young & Fulwiler, 1986). Information gathered from students is attained from case studies, interviews, and anecdotes with additional support gleaned from student satisfaction surveys, pre- and post-tests, and final course evaluations. This research is used to design strategic plans and develop models, methods, and applications. Only recently have leaders such as Chris Anson (2004), renowned researcher and theorist in the development of WAC and WID models, called for rigorous evaluation of the relationship between writing to learn and 1) content knowledge; 2) intellectual development; 3) better disciplinary writing.

The relation between writing and learning seems to call for a more cognitive study of the writing process, research that has typically been conducted in acontextual situations. The current research paradigm in the field of composition studies, however, has moved away from the cognitive and process-oriented movement to a radically contextualized, situated model. Within this research philosophy, Selfe (1997) insists that assessment move away from a positivistic view and instead treat participants and programs as existing primarily within the specific context. If one can no longer make generalizations about the writing process independent of context, questions like those Anson poses must be placed within post-process research like genre analysis (Bazerman, 1988), rhetorical analysis (Graves, 2005), or activity theory (Russell, 1997). Such investigation still requires careful study of writing and writers while demanding more careful study of the writing activity in social context.

Fortunately, recent discipline-specific education research has enriched our understanding of the situational aspects of learning, allowing us to more readily interpret composition that is embedded within a disciplinary class. Physics education research has revealed common misconceptions, epistemologies, and affect-issues (e.g. gender bias or learning style preferences) that interfere with the learning process. While the student is still seen as an individual learner, the frequency with which these issues appear in statistical studies suggests that it is possible to make some generalizations about the learning process within an introductory physics classroom. It is the knowledge of these situational idiosyncrasies that allows us to address Anson's questions, understanding the cognitive process with a proper respect for the disciplinary situation. In this paper we lay out the details of how such a research agenda may proceed within the context of an introductory science classroom and present evidence for its validity. We therefore lay the foundation for a wide range of subsequent research to evaluate the fundamental claims of WAC and WID.

*Background: Writing analysis and research*
Ackerman (1993) provided an early review of the research regarding WAC and the "writing-to learn claims." In this review, he argued that "writing specialists tend to ignore the second half of the write-to-learn equation 'learning and knowledge' and believe that the process and attributes of writing will inevitably lead to learning." In his review, Ackerman found 35 studies in which researchers attempted to give "empirical scrutiny" to the relation between writing tasks and the promotion of learning but also noted that such studies "excluded the writing of multiple drafts or writing in more complex, social relations" which modeled actual academic writing behavior in the discipline. Ackerman thus urged researchers to study the act of writing as evaluation of and participation in disciplinary knowledge and to use text quality to analyze the cognitive processes associated with writing. After reviewing studies conducted in both the sciences and humanities Ackerman concluded that writing likely does "complicate and thus enrich the thinking process" but "only when writing is situationally supported and valued." That is, writing works as an aid to learning only when integral to disciplinary instruction. For example, student performance on multiple-choice tests did not appear to benefit from writing assignments.

Supporting Ackerman's conclusions, Klein (1999) also noted the lack of empirical measures that demonstrate writing's value to the learning process. Specifically, Klein argued that it is the cognitive processes assumed in models of writing to learn which "have long gone without direct empirical examination" (p. 207). Given the significance of the claim and implications for teaching within a WAC or WID model the various assumptions about the processes of writing and processes of learning require further research both to evaluate the cognitive processes that influence writing to learn and to measure the role of social context.

WAC and WID initiatives make explicit the striking claim that writing assignments facilitate learning of disciplinary content and mandate a role for writing in higher education with implications for every discipline. Given these claims and implications, it is neither surprising nor out of place for educational research from across the disciplines to investigate the relationship of writing to disciplinary content learning. Some of the most interesting, primary and recent educational research has come from the fields of math and science (Ediger, 2006; Ellis 2004; Florence & Yore, 2003; Patterson 2001; Sandoval & Millwood, 2005; Yore, Hand, & Florence 2004). A meta-analysis of 48 school-based writing-to-learn programs (Bangert-Drowns, Hurley, and Wilkinson, 2004) cited previous claims for writing as a strategy for enhancing learning but concluded that the whole body of research offered ambiguous conclusions. At best, a thorough review of research indicated only that "writing does appear to facilitate learning to some degree under some conditions" (Bangert-Drowns, Hurley, and Wilkinson, 2004, p. 32). This research attempts to take up the rather difficult question: "how might writing about subject matter content improve the learning–that is, the retention and understanding–of that content?" by applying statistical analysis to research findings in a comprehensive literature review. While conclusions show some positive effects of writing to learn activities, confirmation from additional primary research is still needed. If in early research design "the nature of beneficial writing tasks, the pedagogical contexts in which they should be embedded, and the ways that learning should be assessed were unidentified" (p. 31), more controlled research environments and multiple research designs that consider how writing to learn operates on student cognition in real classrooms are necessary and, finally, possible.

The research described in this paper attempts to apply education research in the disciplines in a more strategic system to develop the recommendations made by Ackerman, Klein, and Bangert-Drowns et al. to address the questions raised by Anson. This kind of discipline-specific research has the potential to enrich our understanding of how students work and think while in the act of writing within a disciplinary class. In this study we blend three research activities: 1) primary trait scoring, 2) revision-based analysis, and 3) key stroke analysis. These assessment strategies represent both traditional and contemporary strategies used in composition research and are here applied in a multimodal research design to analyze the exchange & structure, expression & arrangement, and understanding of information.

As a research instrument primary trait analysis (PTA) establishes learning outcomes as criteria for assignment-based assessment. Identified criteria may be assignment-specific or consistent throughout a class. In its initial formulations (Lloyd-Jones, 1977) PTA focused on the specific approach that a writer might take to be successful on a specific writing task, for example a physics lab report. PTA scales are useful beyond the individual written work and beyond program assessment; primary trait rubrics remain the major approach to assessment of student mastery over rhetorical purposes. The primary traits to be analyzed can be as general as "organization of ideas" or as specific as the particular rhetorical moves in a research paper (e.g. Swales, 1987).

Revision is commonly regarded as a central part of the writing process, in part because it enhances the final written document and also because it requires students to rework ideas, thus potentially enhancing learning (Flower & Hayes, 1981). Faigley and Witte (1981) recognized the need to analyze revision in relation to text by distinguishing between text-based revisions that affect meaning and surface changes that do not. Faigley and Witte were mainly concerned with elements related to linguistic operations. Schwartz (1983) added to these observations by observing three major revision patterns: (1) language regeneration, (2) structural reformulation, and (3) content reassessment. Schwartz's model is largely concerned with the writer's attention to content, language, and mechanics. Berkenkotter (1983) added attention to audience as a major concern for revision, studying the writer's personal composing habits, mode, and genre.

The ability to record a writing episode has become increasingly more sophisticated and, importantly, less intrusive. Recently, methods to capture each keystroke as an essay is developed (on a computer) have been developed, as have associated techniques to analyze this data (Kolberg, 1998). These tools, most prominently S-notation, highlight the order in which revisions occur within a single document and represent the order, range and internal structure of writing and revising activities. Of fundamental importance is the non-invasive nature of this method. Students can compose a document wherever they choose in a variety of editors on their own computers (we have also developed an on-line version). It is therefore possible to conduct empirical studies of revision strategies without intruding on the writing process.

While these methods are not new, none have been used together within a disciplinary context to assess the relationship of writing to learning. Each method accomplishes a separate task. Primary trait analysis captures the generic merit of the final written document as a sample of disciplinary discourse. Version-control captures revisions made between writing sessions, allowing us to note general changes in a written document and potential shifts in content knowledge. Key-stroke analysis captures revisions made during a single writing session, thus emphasizing learning as it might occur during the actual writing process. By applying these three methods in a multimodal engagement and to a disciplinary set of writing practices this study presents alternatives to student assessments or program evaluations as research into claims of WAC and WID. In so doing, this study works as a unique model for sustained, empirical and rigorous inquiry of writing-to-learn.

*Experiment Context: Explorations in Physics*

The context for our research is *Explorations in Physics (EiP)*, an introductory physics course designed for non-science majors.  *EiP* is a concept-oriented physics course that is based upon physics education research and incorporates guided-inquiry techniques with small group projects in a collaborative learning environment. *EiP* centers on common themes that run through the different sciences with an emphasis on physical science.  At Rochester Institute of Technology, *EiP* draws its students from a variety of majors including information technology, computer science and economics.  As these students' careers will require the ability to learn and evaluate new information and to communicate this information to a varied audience, *EiP* seeks to instill in students a level of scientific literacy.  A distinguishing feature of *EiP* is the emphasis on student-directed projects, on which students spend one-third of the quarter.  Student groups choose a topic, write a joint project proposal, and work together to design and conduct relevant experiments.  Projects culminate with a written report as well as a poster presentation.  The remaining two-thirds of the course is spent working through two activity guides, which guide the students through experiments on different topics.  The division of time between projects and guided inquiry is relevant for this study since we note differences in student writing between project and other reports.

*Primary Trait Analysis*

Primary trait analysis is a procedure to categorize the assignment-specific traits that are observable criteria for particular tasks in different written documents. The analysis begins by identifying the major aspect(s) or criteria that readers consider when assessing the written product or behavior.  As such, it's value lies in a reliable, quantitative assessment of student writing.  It does not provide direct insight into student's composition process, and it is a seductive mistake to make inferences about what a student knows based solely on primary trait analysis.

We developed a primary trait rubric to apply to student lab reports.  The role of a lab report is complex.  Students are expected (and instructed) to provide some background, a detailed yet succinct description of the activities, and then place their findings within a larger framework.  It is not surprising that, when faced with such a far reaching project, students often struggle to develop a coherent essay.  Based upon analysis of student papers, criteria were devised to code sentences as **M**otivation, **P**rocedure, **O**bservation, **I**nference, or **S**peculation (*MPOIS*).  This trait analysis, while extremely simplistic, captures many elements of scientific writing including (a) the importance of carefully describing one's procedure, (b) the primacy of observation, and (c) the difference between inference and speculation.  A sample piece of coded writing is shown in Fig. 1.

> This being found, we began **M** to experiment more with buoyant force and its relation to water displaced by an object when it's dropped in water.  By performing **P** several experiments with an overflow can and measuring the amount of water displaced, we were **I** able to conclude that buoyant force on an object and the amount of water displaced by the same object are related to each other, and seemed to be equal to each other.  Thus, we began **P** to experiment with the possibility of calculating buoyant force on object simply by measuring its volume and predicting the amount of water it will displace.  We created a set standard based on the amount of water displaced by an object of a certain volume, and then used that standard to predict the buoyant force on objects of different volumes.

**Fig. 1:** Writing sample that has been coded with the *MPOIS* (**M**otivation, **P**rocedure, **O**bservation, **I**nference, or **S**peculation) rubric.  It is possible to identify gross features of student writing, such as the number of procedural or inference statements.

We should emphasize that this categorization, especially the difference between inference and speculation, is fluid.  Over the course of the quarter, as a result of the ongoing discussions with students, we gradually settled into a reliable articulation.  While students had little trouble with identifying statements as motivating (anything providing background or motivation for research. i.e. "It is interesting that boats float while rocks sink.") or procedural (i.e. "We placed the block in the water."), the remaining three criteria required some discussion.  To our surprise, it became apparent that many students did not know what was meant by an observation. This was illustrated in a classroom discussion about the following sentence on buoyancy:

> While dropping the wooden blocks into water, the buoyant force proved to be a rather important factor pushing against the object.

While some students claimed that this was an observation of the buoyant force, others argued that this was an inference, the observation being that the block floated. The class finally agreed (with subtle urging from the faculty) that the observation was that the block floated at the surface of the water. Based on other observations, an appropriate *inference* was that the forces on the block were in balance. The existence of the buoyant force in balance with gravity was then a reasonable *speculation.*

This primary trait analysis was used to analyze approximately 600 student writing samples during the 2003-2004 academic year. Students were required to submit weekly essays that documented the week's activities while placing them within a larger context. Essays were to be written at the level of a typical *Scientific American* article. In order to allow students to incorporate faculty feedback, as well as to encourage the connectedness of all the activities, journals were cumulative. Students were told to revise their initial entry and integrate the new weeks' activities into one seamless essay. During the quarter, students wrote three separate essays (two topical + one project), each of which was revised 3-4 times. Feedback included an assessment of the student's writing ability, including comments on the extent of the student's revisions. Students were given a description of the assessment scheme and numerous writing samples that demonstrated both positive and negative elements of scientific writing

Figure 2 shows the frequency of occurrence from two different journals, *A* and *B*. (*A1/B1* is the first draft of journal *A/B*, *A2/B2* the last.) Several features are noteworthy. Speculative sentences, which initially make up 35% of all student writing drop after feedback to about 25% and remain at that level for the rest of the quarter. Motivating statements, a catch-all category that included unnecessary fluff, also fell by about 10%. Inference seems to be the most difficult type of statement to impact. Despite repeated attempts to define an inference and give examples, students never understood this well enough to incorporate them into the essays.

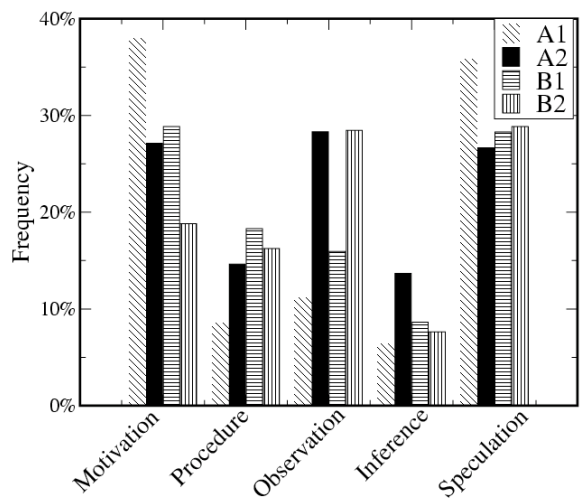

**Fig. 2:** Frequency of occurrence of sentences designated as Motivation, Procedure, Observation, Inference and Speculation in student essays. *A1/A2* refer to the first and last versions of the first student essay, *B1/B2* to the first and last versions of the second student essay. Significant shifts are seen in student style, away from unnecessary motivation and speculation and towards a more complete description of procedure and observations. Salient inferences, unfortunately, do not show a significant increase frequency.

One would expect that the experience of writing journal *A* (with its four revisions) would have some impact on the second journal. One can see in Fig. 2 how the first draft of the second journal (*B1*) has fewer motivation and speculation sentences than *A1*. The emphasis on procedure also remains, and so we believe is justified to claim that some shift in style has occurred. The low frequency of observation-type sentences in both journals *A1* and *B1* is due, we believe, to the fact that the first week's activities involved few direct observations rather than a regression on the part of the students.

Coding the primary trait criteria allows us to look at complex performances and to identify precise outcomes. The coding makes possible a graphical display of student writing; the *writing graph* for the full journal from which Fig. 1 was taken is shown in Fig. 3. Note the large number of statements classified as observations and how frequently an observation is followed by an inference. Twenty-two of the twenty-five inference and speculative statements (88%) are immediately preceded by an observation or description of procedure. This could indicate a favorable epistemology that recognizes that the foundations of physics principles lie in the observations, as opposed to common belief that observations exist to support ideas. It is impossible, however, to determine the truthfulness of this claim without interviewing the student.

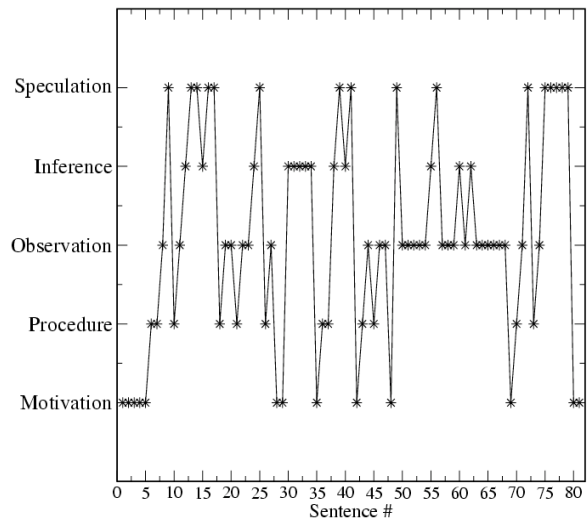

**Fig. 3:** Writing graph of a student writer's paper. Note the preponderance of statements that describe an observation and that, with few exceptions, inferences and speculations are preceded by an observation or description of procedure. This may indicate a favorable epistemology, the student recognizing that physics ideas must be supported by the observations.

The writing graph for a second student writer is shown in Fig 4; the differences are immediately apparent. There is a lack of inference and speculative sentences, a dearth particularly noteworthy from the concluding sentences where one might expect more summative types of sentences. Also apparent is a "procedure-observation" cycle (sentences 21-26). This is a common finding in science writing where students write "We did this and saw this" without making any broader conclusions.

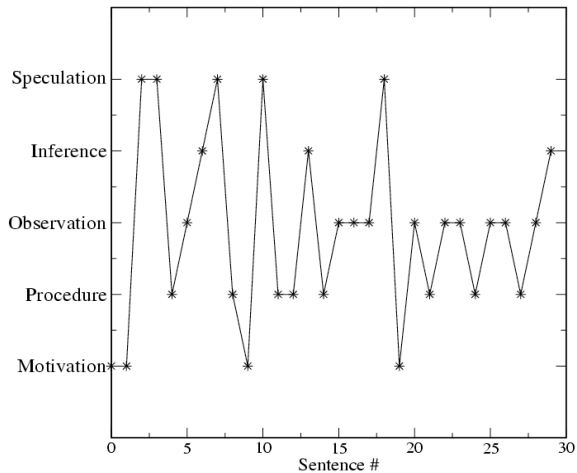

**Fig. 4:** Writing graph of an second student writer's paper. Note the dearth of speculative or inferential statements, an absence particularly significant at the end of the essay. The "Procedure-Observation" cycle is one quite common in introductory science writing.

Primary Trait Analysis is useful because it defines some of the important characteristics of discipline-specific writing. The *MPOIS* rubric is appropriate for use in physics because the inclusion of criteria and the sequencing of criteria in lab reports guide students through the epistemological assumptions that support study of physical phenomena. This particular rubric is designed to fit the *EiP* course context. It garnered immediate endorsement as an instructional tool from physics faculty untrained in generic composition, indicating its value within the physics discipline. Subsequent use of this PTA (and similar rubrics in different disciplines) can add to our understanding about the connection between writing activities and discipline-based epistemologies and content understanding.

*Revision-based Assessment*

It is generally accepted that for writing to be pedagogically effective the student must spend a large fraction of the time revising, either explicitly on paper or implicitly in his/her head (Flower & Hayes, 1981). If a student sees the purpose of a writing assignment as his/her opportunity to display his/her knowledge to the instructor, then s/he proceeds to write down what s/he knows and then stops. There is no need for revision. In fact, many students appear incapable of revising even when explicitly instructed to do so, tending to "revise" by excising anything the instructor has indicated does not belong without replacing it with necessary text to hold the document together. Or, if the instructor notes that a particular turn of phrase or idea is valuable, the student then tries to repeat this as often as possible. Either way, the student is not seeing revision as an opportunity to reorganize his or her thoughts. And yet, it is precisely this mental reorganization that may lead to real content learning.

In order to attempt to lead students into a reflective writing mode that encourages revision, we implemented two-part *revision-based writing assignments* in the *Explorations in Physics* classroom. In the first part of the assignment, students are asked to write an essay with a well-defined length. Once this is turned in students are immediately, and without any feedback, asked to revise their essay to *half its original length*. This drastic change ensures that simply cutting extraneous words cannot reach the word limit. A total of 60 students (30 per quarter) completed 4 assignment pairs. A sample pair prompt is

> Write a **250 word essay** on the connection between net force and how objects act when placed in a liquid (or released within the fluid). You should include *your own class observations* to support your reasoning.

which is followed by

> Take your essay from HW #1 and re-write it, shortening it to **125** words.

Students submit these assignments in electronic form using a word processor with the track changes feature enabled. This allows the instructor or researcher to quickly see the differences between the first and second versions. On a computer screen inserted and deleted text is colored differently; here text inserted in the second version is italicized while excised text is crossed out. A sample of student work is shown in Fig. 5.

> Sinking and floating usually seems like a simple idea. ~~The activity that involved using force meters is the one that really opened my eyes to the world of sinkers and floaters. Even though this activity had nothing to do with water, the concepts finally fit together in my mind.~~ Obviously, gravity ~~is present in pushing~~ *pushes something down* ~~towards the earth~~ when it is dropped. ~~in water. It was also obvious that there was another force pushing up on the floaters to keep them at the surface,~~ *There must also be another force*

> *from the water pushing on the object,* ~~but~~ *yet this force* ~~must be~~ is in equilibrium with gravity or the object would *always move.* ~~simply fly up out of the water.~~ We ~~would~~ call this a ~~net force of zero~~ *zero net force, since the forces* ~~were~~ *are* ~~in~~ balanced. The ~~upward~~ force from the water ~~that I mentioned was really the one that we had not understood previously to the activity~~ *is called buoyancy.* ~~However, using the force meters and lifting weights slightly with our hands, but leaving enough hanging that the force did not drop to zero, I finally understood the idea of this ``buoyant force'' that is present in the water.~~ While the water does not ~~not fully~~ support the object *entirely, it presses up on it* ~~just~~ enough to create an ~~sort of~~ ``apparent weight'', though ~~Of course,~~ *the object's true weight* ~~still does not change~~ *never changes.* ~~Yet, we have all noticed that some things feel lighter in water than they do out of water.~~ *By* ~~taking this apparent weight and subtracting it from its real weight~~ *subtracting these two weights, we* ~~should be able to see~~ *can find* ~~how much force the water is using to push up on the object we drop into it~~ *the buoyant force.* ~~Once these forces eventually balance out, we end up with a net force of zero, causing the object to settle somewhere in the water.~~ What decides if the object sinks or floats is where ~~these forces balance out~~ *in the water the zero net force occurs.*

**Fig. 5:** Student essay on the relationship between net force and buoyancy, as prompted by the assignment described in the text. The student initially wrote a 250 word essay and then revised it to 125 words. Because the **track changes** feature of Microsoft Word was enabled, we can clearly see where the student made changes. *Italicized* text has been inserted during the revision process while smaller crossed out text has been deleted.

One can see significant changes between versions. Rhetorically, the writing has become more active ("gravity pushes something down", "We call this a zero net force") and transitions are more sophisticated ("There must also be" and "By subtracting these two weights"). From a physics perspective, the student is now focusing on the underlying concept of net force rather than the apparent differences between floating and sinking. While it has often been posited that improved writing is indicative of greater content understanding, we believe this is the first instance where a shift in writing style has been directly connected with an evolution in understanding. This type of revision, seen in many students in our study, is powerful evidence that the *writing process itself* can bring about a change in understanding, i.e. actual learning.

We have found the draconian shortening is necessary to impress upon students the need to reorganize their essays. Simply asking students to revise or slightly shorten an essay results in "revision by excision": students cut extraneous words but do not significantly revise the content. This avoids the very process we desire students to perform: the development of a coherent framework that structures the essay. An example of this is shown in Fig. 6.

> ~~Yes there is a connection because the forces determine how it moves through the liquid~~ With a small ball of clay tested in a jar of water to see if it would float. It ~~immediately~~ sunk to the bottom showing that the net downward force had to be stronger coming down than going up. ~~The next object that we experimented on was a large bouncy ball that just floated on the top of the water. This showed us that the net force was zero because it was just floating on the water causing equilibrium between upward and downward forces.~~ The last object that we used ~~in our experiment~~ was a ping pong ball. We took the ping pong ball and submerged it under water and it ~~immediately rose straight up~~ rose to the top. This showed that the upward ~~net~~ forces were much stronger than the downward net forces. There were two forces acting up on the ping pong ball which were buoyant force and the forces of the water pushing up from below. ~~One force was acting down on the ping pong ball and that was the force of the water acting down from above.~~ Because the ping pong ball was going up we know that the forces acting upward had to be greater than the forces acting downward.

**Fig. 6:** Student essay buoyancy that shows significant excision but little thoughtful revision. The student was, in fact, unable to reduce the length to the requested 125 words, in part due to the fact that the first 2 sentences are unrelated to the next 5.

Note the complete absence of any inserted text. The student cut as many words as he felt he could but then became frustrated at being unable to reach the desired length of 125 words. This prompted instruction on *coherence*. It was observed that his introductory and concluding sentences were unrelated and that the first two sentences could be, but were not, thematically related to his last five.  With this explicit guidance toward writing a coherent essay the student was finally able to reach the desired length with an essay that indicated a much improved understanding of the physical phenomena of net force and buoyancy *despite the fact that no explicit content instruction had occurred*.

A natural question is whether, having undergone this revision exercise, students are able to generate better first drafts.  That is, can they now revise *implicitly* while composing the first draft.  To investigate this, students were given asked to write a 125 word essay as a first draft rather than a revision of a longer essay.  To simplify the task, the assignment prompt was more specific than previous prompts:

> 60 milliliters of a mystery substance weighs 90 ounces.  Write a 125 word (no more!) essay explaining, in detail, how you can use this information to predict the weight of 90 milliliters of the substance.  Include in your essay a description of how this process can be related to the idea of density.  *Pay close attention to writing a meaningful introductory and concluding sentence, as well as the transitions between sentences.*

A sample essay that resulted is shown in Fig. 7, and is typical of what was turned it.  Note the absence of any transition that would build a sense of coherence.  The rhetorical device that connects the introductory and concluding sentences is the extremely simplistic repetition of the phrase "Its (sic) all about the...", which is itself ambiguous.  Our students, therefore, did not show evidence of learning how to self-revise and write reflectively without an assignment that forced them to.

---

Its (sic) all about the ounces per milliliter.  We take 90 ounces and divide it by 60 milliliters.  This gives us the weight of 1 milliliter of the substance, which is 1.5 ounces. We multiply this by the number of milliliters that we wish to know the weight of, 90 milliliters, to get the final weight, 135 ounces. These are the same steps used to determine the density of an object. Density is the measurement of how much of a substance occupies a volume of space. The measurement of 1.5 ounces is also how much of this substance is present in 1 milliliter of volume.  Its (sic) all about the mass per unit of volume.

---

**Figure 7:**  Sample first-draft short essay.  The paragraph shows little of the coherence seen in second-drafts indicating that students do revise unless explicitly forced to by a paired assignment.

If learning is an evolution of understanding, then it should be reflected not only in what is written, but what is changed. Revision analysis is a tool that allows us to quickly identify the key changes in a document that might indicate learning has occurred. The evidence presented here is not intended as definitive proof of learning in the *EiP* classroom. Rather, it highlights the possibilities of a research agenda that focuses on revision both from a rhetorical and content perspective. The explicit connection between the two is highly suggestive and provides a clear path for future research.

*Key-stroke analysis*

While the version control software shows differences between first and second drafts, key-capture software allows us to see how a specific draft is written (or revised). A format, labeled S-notation (\cite{Kollberg96, Severinson96}), has been developed to facilitate the reconstruction of a writing episode. In S-notation, interruptions in writing are noted by $\|_i$ where i represents the order of interruptions. Insertions and deletions are marked with {inserted text}$^i$ and (deleted text)$^i$. An example of this notation (taken directly from \cite{Kollberg96}) is

I am writing a {short}$^1$ text.$\|_1$. It will (probably)$^2\|_3$ be revised (somewhat)$^3$ later.$\|_2$. Now (I am$\|_4$)$^4$ it is finished.

The interpretation of this is as follows: First the subject wrote "I am writing a text." Then they went back and inserted the word short. Next they wrote "It will probably be revised somewhat later", after which they deleted the words "probably" and "somewhat." Finally, they wrote "Now I am", changed their mind and deleted the words "I am" and inserted "it is finished" to produce:

*I am writing a short text. It will be revised later. Now it is finished.*

A more substantial example of S-notation, shown in Fig. 8, illustrates the type of significant revision event that we believe indicates the potential pedagogical value of writing assignments. This particular sample was taken as part of a pilot project; the student was not enrolled in Explorations in Physics, but was of comparable educational background. While in this case the student was aware of the key-capture software, we have also developed a stand alone and web-based version that students can use transparently throughout the quarter. In this case, the student was given the following prompt:

> 60 ml of a mystery substance weighs 90 oz. Write a 250 word essay explaining, in detail, how you can use this information to predict the weight of 90 ml of the substance in two ways: through the ideas of proportionality and density.

In response, the student wrote the essay shown in Fig. 8.

---

{Density is an important part of modern science, finding its way into many fields including chemistry, optics, and physics. Typically, if given a total amount of substance and a weight, one can find the weight of a different amount of the same substance. One may also be able to use formulas to calculate a density PAUSE by plugging numbers into it, but what it all boils down to, in this case, is the total amount of "stuff" within a specific volume.}[4]||[5]

(I don't know anything about density; however,)[3] ||[4] There are at least a couple ways of determining {it}[7] {the weight of a substance} (, or at least I'm told)[5]||[6] {when its mass and volume are known}[6] ||[7]. PAUSE The simplest way of determining it would be through the use of proportionality. In the example, a sample of 60 ml of substance weighed 90 ounces. The question asks to determine how much 90 ml of the same substance would weigh. To set this (this up||[1])[1] proportionality one must PAUSE set up a ratio. So, 60 ml over 90 ounces is equal to 90 ml over x ounces. By cross-multiplying and (dividing||[2])[2] solving for x, one can find the value of x, which is the PAUSE weight of the 90 ml of substance. (Another way of determining the total amount of mass)[9]||[3], PAUSE Another way of determining the total weight of the system would be to simply divide PAUSE 90 ounces by 60 ml. Density is defined as mass over volume, so by doing this, one has found the density. Once the density has been found, a person simply must multiply this number by a new given amount to find the new mass. This is the same procedure as before, but practiced in a different way.

---

**Figure 8:** Writing sample rendered using S-notation. The enclosure of the introductory paragraph in braces {} indicates it was added after much of the following text was written. Similarly, the phrase in parentheses (I don't know anything about density) was the first text written and removed just before the new introductory paragraph was written.

While there are many fascinating events that occurred throughout the writing process, here we will focus on one, dealing with the very first words the student wrote: "I don't know anything about density; however, there are at least a couple of ways of determining in, or at least I'm told." He then spends 2.5 minutes writing about how to solve the problem by setting up a simple proportionality. Then, realizing that he now needs to describe a second solution involving density, he pauses and thinks without writing for 30 seconds. His next action is incredibly suggestive: he erases the opening clause "I don't know anything about density" and inserts a placeholder introductory paragraph about density's importance including a definition of density as "the total amount of stuff in a volume." Having thus concluded that he *does* know something about density he returns to the end of his essay, describes how one could use density to answer the question, re-defines density as mass per volume, and, remarkably, relates this to his previous use of proportionality, writing that "this is the same procedure as before, but practiced in a different way."

How can one interpret this writing process? The explicit statement ("I don't know anything about density") is a candid self-reporting of the student's initial state of ignorance. As the student writes, it appears that he reaches a series of increasingly sophisticated conceptions of density. He begins with the standard, oft-memorized formula of "mass over volume." The relative crudeness of his understanding is seen in his use of the term "over" for the mathematical operation of division and by his earlier statement that one "may also be able to use formulas to calculate a density by plugging numbers into it." Clearly he is seeing density as a relation between given numbers (an object's weight and volume) and not as an abstract characteristic of a substance. In between writing these two sentences, however, he gives a clue to a more sophisticated view when he writes that "it all boils down to, in this case, is the total amount of 'stuff' within a specific volume." It is in his concluding sentence, however, that he appears to make the most significant leap of understanding. By relating density to his previous method of proportionality he unifies two concepts that, when he sat down to write, were completely separate in his mind. It is this synthesis that is at the heart of writing coherently and it is particularly impressive that the student accomplishes this without *any* instruction. The writing task, it seems, forced him to teach himself about density.

Our goal in presenting this case study is to demonstrate both the possibility and promise of probing the writing process within a disciplinary context. The results strongly suggest an evolution in the student's understanding within a single writing episode. Regardless of the details of this learning (e.g. permanency, completeness, etc.), we are now closer to connecting explicitly discipline-specific content understanding with the writing process. While it would be foolhardy to draw general conclusions about learning from a single case study, key-capturing presents a significant, new, and viable research method to track both writing and content understanding within the disciplinary situation.

*Conclusion*

We have argued for a research agenda on writing-to-learn (WTL) based upon a multi-modal approach within a disciplinary context. This directly tests the fundamental assumption of WAC and WID programs: writing tasks associated with course content can and do improve the quality of students' overall learning of subject matter. Our approach is a radical departure from prior assessments of WTL, which have traditionally used general writing tasks (e.g. short answer summaries, note-taking, and journaling) with uncertain results. Our agenda is consistent with recent theory that learning-to-write, and thus writing-to-learn, is not a general activity but one thoroughly embedded in disciplinarity.

Most recent research agendas associated with WAC and WID programs call for assessing writing and learning in context but have struggled with the implementation of this assessment. Because writing is intrinsically personal, standard approaches such as "speak-aloud" interview sessions necessarily intrude on the student's cognitive process, the very phenomena under study. One significant advance presented in this paper is the incorporation of standard research methods–primary-trait analysis, revision-based assessments, and key-stroke analysis–into the classroom in a manner less intrusive to the writing process.

In addition to establishing the viability of embedding these research methods in the classroom, we have also shown the methods' enormous potential at establishing a connection between writing and learning. In particular, we are now poised to evaluate several cognitive theories about writing-to-learn within the disciplinary context. The multi-modality of the agenda is critical to this point. This type of triangulation (Creswell, 1998; Yin, 2003) is a well-established technique in qualitative and case-study research. Having the evolution in writing as assessed by a primary-trait analysis reinforced by revision- and key-stroke-analysis strengthens any conclusions that emerge from this research.

While the particulars of each assessment are disciplinary-based, this broad research agenda is not necessarily tied to our chosen context; rather, it is immediately ready for adaptation in almost any disciplinary course with a significant writing component. The groundwork is therefore prepared for large-scale research on writing-to-learn in many different disciplinary situations.

science writing practices. *Journal of Research in Science Teaching 41*(4), 338-369.

Walvoord, B.E. (1992). Getting started. In S.H. McLeod and M. Soven (Eds.). *Writing across the curriculum: A guide to developing programs*. Newbury Park, CA: Sage.